\documentclass{acm_proc_article-sp}

\usepackage{times}

\usepackage{graphicx}%
\usepackage{subfigure}%

\usepackage{xspace}%
\usepackage{listings}%
\usepackage{algorithm}
\usepackage{algorithmic}

\usepackage{url} 

\newcommand{\CREATE}{\mbox{\lstinline$CREATE$}\xspace}
\newcommand{\READ}{\mbox{\lstinline$READ$}\xspace}
\newcommand{\WRITE}{\mbox{\lstinline$WRITE$}\xspace}
\newcommand{\APPEND}{\mbox{\lstinline$APPEND$}\xspace}
\newcommand{\SYNC}{\mbox{\lstinline$SYNC$}\xspace}

\def \setmyId $Id: #1${\gdef\myId{#1}}
\setmyId $Id: paper.tex 3778 2008-12-19 16:44:17Z gantoniu $

\begin{document}

\title{%
  BlobSeer: How to Enable Efficient Versioning for Large Object
  Storage under Heavy Access Concurrency
%
}

\toappear{Permission to make digital or hard copies of all or part of 
this work for personal or classroom use is granted without fee provided 
that copies are not made or distributed for profit or commercial advantage 
and that copies bear this notice and the full citation on the first page. 
To copy otherwise, to republish, to post on servers or to redistribute to lists, 
requires prior specific permission and/or a fee. 
\par
{\confname DAMAP 2009}, March 22, 2009, Saint Petersburg, Russia.\par
Copyright 2009 ACM 978-1-60558-650-2\ ...\$5.00} 

\numberofauthors{3} 
%
\author{
%
%
\alignauthor
Bogdan Nicolae
    \\\affaddr{University of Rennes~1, IRISA, Rennes, France}
\alignauthor
Gabriel Antoniu
    \\\affaddr{INRIA, Centre Rennes - Bretagne Atlantique, IRISA, Rennes, France}
\alignauthor  Luc Boug\'e
    \\\affaddr{ENS Cachan/Brittany, IRISA, France}
}

\maketitle

\begin{abstract}
To accommodate the needs of large-scale distributed P2P systems,
scalable data management strategies are required, allowing
applications to efficiently cope with continuously growing, highly
distributed data. This paper addresses the problem of efficiently
storing and accessing very large binary data objects (blobs).  It
proposes an efficient versioning scheme allowing a large number of
clients to concurrently \emph{read, write and append} data to huge
blobs that are fragmented and distributed at a very large
scale. Scalability under heavy concurrency is achieved thanks to an
original metadata scheme, based on a distributed segment tree built on
top of a Distributed Hash Table (DHT). Our approach has been
implemented and experimented within our \emph{BlobSeer} prototype on
the Grid'5000 testbed, using up to 175 nodes.
\end{abstract}

\section{Introduction}
\label{sec:intro}

Peer-to-peer (P2P) systems have extensively been studied during the
last years as a means to achieve very large scale scalability for
services and applications. This scalability is generally obtained
through software architectures based on autonomic peers which may take
part in a collaborative work process in a dynamic way: they may join or
leave at any time, publish resources or use resources made available
by other peers. P2P environments typically need scalable data
management schemes able to cope with a growing number of clients and
with a continuously growing data, (e.g. data streams),
while supporting a dynamic and highly concurrent environment.

As the usage of the P2P approach extends to more and more application
classes, the storage requirements for such a large scale are becoming
increasingly complex due to the rate, scale and variety of data. In
this context, storing, accessing and processing \emph{very large,
unstructured data} is of utmost importance. Unstructured data
consists of free-form text such as word processing documents, e-mail,
Web pages, text files, sources that contain natural language text,
images, audio and video streams to name a few. 

Studies show more than 80\%~\cite{Gri08Unstructured} of data globally
in circulation is unstructured. On the other hand, data sizes increase
at a dramatic level: for example, medical experiments~\cite{1321583}
have an average requirement of 1~TB per week.  Large repositories for
data analysis programs, data streams generated and updated by
continuously running applications, data archives are just a few
examples of contexts where unstructured data that easily reaches the
order of 1~TB.

Unstructured data are often stored as a \emph{binary large object
  (blob)} within a database or a file. However, these approaches can
hardly cope with blobs which grow to huge sizes. To address this
issue, specialized abstractions like MapReduce~\cite{1327492} and
Pig-Latin~\cite{1376726} propose high-level data processing frameworks
intended to hide the details of parallelization from the user. Such
platforms are implemented on top of huge object storage and target
high performance by optimizing the parallel execution of the
computation. This leads to \emph{heavy access concurrency} to the
blobs, thus the need for the storage layer to offer support in this
sense. Parallel and distributed file system also consider using
objects for low-level storage~\cite{1362659,WeilCeph06,945450}. 
In other scenarios, huge blobs need to be used concurrently at the highest 
level layers of applications directly: high-energy physics applications, 
multimedia processing~\cite{CaseyKurth07Multimedia} or
astronomy~\cite{NicAntBouCLUSTER2008}.

In this paper we address the problem of storing and efficiently
accessing very large unstructured data
objects~\cite{MesnierOSD03,1321583} in a distributed environment. We
focus on the case where data is \emph{mutable} and potentially
accessed by a very large number of concurrent, distributed processes,
as it is typically the case in a P2P system. In this context,
\emph{versioning} is an important feature. Not only it allows to roll
back data changes when desired, but it also enables cheap branching
(possibly recursively): the same computation may proceed independently
on different versions of the blob. Versioning should obviously not
significantly impact access performance to the object, given that
objects are under constant heavy access concurrency.  On the other
hand, versioning leads to increased storage space usage and becomes a
major concern when the data size itself is huge. Versioning efficiency
thus refers to both access performance under heavy load and reasonably
acceptable overhead of storage space.

Related work has been carried out in the area of parallel and
distributed file systems~\cite{Lustre,PVFS,945450} and archiving
systems~\cite{DeepStore}: in all these systems the metadata management
is centralized and mainly optimized for data reading and appending. In
contrast, we rely on metadata decentralization, in order to introduce
an \emph{efficient versioning} scheme for huge, large-scale
distributed blobs that are concurrently accessed by an arbitrarily
large number of clients which may read, write or append data to
blobs. Our algorithm guarantees atomicity while still attaining good
data access performance. Our approach splits a huge blob into small
fixed-sized pages that are scattered across commodity data
providers. Rather than updating the current pages, completely new
pages are generated when clients request data modifications. The
corresponding metadata is ``weaved'' with old metadata in such way as
to offer a complete virtual view of both the past version and the
current version of the blob. Metadata is organized as a segment-tree
like structure (see Section~\ref{sec:metadata}) and is also scattered
across the system using a Distributed Hash Table (DHT). Distributing
data and metadata not only enables high performance through parallel,
direct access I/O paths, but also favors efficient use of storage
space: although a full virtual view of all past versions of the blob
is offered, real space is consumed only by the newly generated pages.

Our approach has been implemented and experimented within our
prototype, called \emph{BlobSeer}: a binary large object management
service. In previous work~\cite{NicAntBouCLUSTER2008,
  NicAntBouHPDGrid2008} we have handled versioning in a static way:
blobs were considered huge storage objects of predefined, \emph{fixed}
sizes that are first allocated, then manipulated by reading and
writing parts of them. However, in most real life scenarios, blobs
need to dynamically grow, as new data is continuously gathered.  This
paper improves on our previous work as follows. First, we introduce
support for dynamic blob expansion through atomic append
operations. Second, we introduce cheap branching, allowing a blob to
evolve in multiple, completely different ways through writes and
appends starting from a particular snapshot version. This may be very
useful for exploring alternative data processing algorithms starting
from the same blob version.

The paper is organized as follows. Section~\ref{sec:specs} restates
the specification of the problem in a more formal
way. Section~\ref{sec:design} provides an overview of our design and
precisely describes how data access operations are handled. The
algorithms used for metadata management are discussed in
Section~\ref{sec:metadata}. Section~\ref{sec:impl} provides a few
implementation details and reports on the experimental evaluation
performed on multi-site grid testbed. On-going and future work is
discussed in Section~\ref{sec:conclusion}.


\section{Specification}
\label{sec:specs}

Our goal is to enable efficient versioning of blobs in a highly
concurrent environment. In such a context, an arbitrarily large number
of $n$ clients compete to read and update the blob. A blob grows as
clients append new data and its contents may be modified by partial or
total overwriting.

Each time the blob gets updated, a new snapshot reflecting the changes
and labeled with an incremental version is generated, rather than
overwriting any existing data.  This allows access to all past
versions of the blob. In its initial state, we assume any blob is
considered empty (its size is 0) and is labeled with version 0.(Note
that our previous work~\cite{NicAntBouCLUSTER2008,
  NicAntBouHPDGrid2008} was relying on different assumptions: the blob
size was statically specified at the initialization time and could not
be extended.)

Updates are totally ordered: if a snapshot is labeled by version $k$,
then its content reflects the successive application of all updates
$1..k - 1$ on the initial empty snapshot in this order. Thus
generating a new snapshot labeled with version $k$ is semantically
equivalent to applying the update to a copy of the snapshot labeled
with version $k - 1$. As a convention, we will refer to the snapshot
labeled with version $k$ simply by \emph{snapshot $k$} from now on.

\subsection{Interface}

To create a new blob, one must call the \CREATE primitive:
\label{sec:intf}
\[
\mbox{\lstinline$id = CREATE( )$}
\]
This primitive creates the blob and associates to it an empty
snapshot 0. The blob will be identified by its $id$ (the returned
value).  The \emph{id} is guaranteed to be globally unique.
\[
\mbox{\lstinline$vw = WRITE(id, buffer, offset, size)$}
\]
A \WRITE initiates the process of generating a new snapshot of the
blob (identified by $id$) by replacing \emph{size} bytes of the blob
starting at \emph{offset} with the contents of the local
\emph{buffer}.

The \WRITE does not know in advance which snapshot version it will
generate, as the updates are totally ordered and internally managed by
the storage system. However, after the primitive returns, the caller
learns about its assigned snapshot version by consulting the returned
value $vw$. The update will eventually be applied to the snapshot $vw
- 1$, thus effectively generating the snapshot $vw$. This snapshot
version is said to be \emph{published} when it becomes available to
the readers.  Note that the primitive may return before snapshot
version $vw$ is published.  The publication time is unknown, but the
\WRITE is \emph{atomic} in the sense of \cite{1270853}: it appears to
execute instantaneously at some point between its invocation and
completion. Completion in our context refers to the moment in time
when the newly generated snapshot $vw$ is published.

Finally, note that the \WRITE primitive fails if the specified
\emph{offset} is larger than the total size of the snapshot $vw - 1$.
\[
\mbox{\lstinline$va = APPEND(id, buffer, size)$}
\]
\APPEND is a special case of \WRITE, in which the \emph{offset} is
implicitly assumed to be the size of snapshot $va - 1$.
\[
\mbox{\lstinline$READ(id, v, buffer, offset, size)$}
\]
A \READ results in replacing the contents of the local \emph{buffer}
with \emph{size} bytes from the snapshot version $v$ of the blob $id$,
starting at \emph{offset}, if $v$ has already been published.  If
\emph{v} has not yet been published, the read fails.  A read fails
also if the total size of the snapshot \emph{v} is smaller than
\emph{offset + size}.


Note that the caller of the \READ primitive must be able to learn
about the new versions that are published in the system in order to
provide a meaningful value for the $v$ argument. The blob size
corresponding to snapshot $v$ is also required, to enable valid s
from the blob to be read.  The following primitives are therefore
provided:
\[
\mbox{\lstinline$v = GET_RECENT(id)$}
\]
This primitive returns a recently published version blob $id$. The
system guarantees that $v \geq max(v_k)$, for all snapshot versions
$v_k$ published before the call.
\[
\mbox{\lstinline$size = GET_SIZE(id, v)$}
\]
This primitive returns the size of the blob snapshot corresponding to
version \emph{v} of the blob identified by $id$. The primitive fails
if $v$ has not been published yet.

Since \WRITE and \APPEND may return before the corresponding snapshot
version is published, a subsequent \READ attempted by the same client
on the very same snapshot version may fail. However, it is desirable
to be able to provide support for ``read your writes'' consistency.
For this purpose, the following primitive is added:
\[
\mbox{\lstinline$SYNC(id, v)$}
\]
The caller of \SYNC blocks until snapshot $v$ of blob $id$ is
published.


Our system also introduces support for branching, to allow alternative
evolutions of the blob through \WRITE and \APPEND starting from a
specified version.
\[
\mbox{\lstinline$bid = BRANCH(id, v)$}
\]
This primitive virtually duplicates the blob identified by $id$ by
creating a new blob identified by $bid$. This new blob is identical to
the original blob in every snapshot up to (and including) $v$. The
first \WRITE or \APPEND on the blob $bid$ will generate a new snapshot
$v + 1$ for blob $bid$. The primitive fails if version $v$ of the blob
identified by $id$ has not been published yet.

\subsection{Usage scenario}

Let us consider a simple motivating scenario illustrating the use of
our proposed interface. A digital processing company offers online
picture enhancing services for a wide user audience. Users upload
their picture, select a desired filter, such as sharpen and download
their picture back. Most pictures taken with a modern camera include some metadata in
their header, describing attributes like camera type, shutter speed,
ambient light levels, etc.  Thousands of users upload pictures every
day, and the company would like to analyze these pictures for
statistical purposes. For example it might be interesting to find out
the average contrast quality for each camera type.

One option to address this problem would be to store the pictures in a
huge database and perform some query when needed. Unfortunately, pictures
are unstructured data: metadata is not standardized and may differ
from one camera brand to another.  Thus, no consistent schema can be
designed for query optimization. Moreover, it is unfeasible to store
variable binary data in a database, because database systems are
usually fine-tuned for fixed-sized records.

Let us now consider using a virtually unique (but physically
distributed) blob for the whole dataset. Pictures are \APPEND'ed
concurrently to the blob from multiple sites serving the users, while
a recent version of the blob is processed at regular intervals: a set
of workers \READ disjoint parts of the blob, identify the set of
pictures contained in their assigned part, extract from each picture
the camera type and compute a contrast quality coefficient, and
finally aggregate the contrast quality for each camera type. This type
of computation fits in the class of map-reduce applications. The map
phase generates a set of (key, value) pairs from the blob, while the
reduce phase computes some aggregation function over all values
corresponding to the same key. In our example the keys correspond to
camera types.

Many times during a map phase it may be necessary to overwrite parts
of the blob. For example, a complex image processing was necessary for
some pictures and overwriting the picture with its processed version
saves computation time when processing future blob versions. Surely, a
map with an idempotent reduce reaches the same result with no need to
write, but at the cost of creating an output that duplicates the blob,
which means an unacceptable loss of storage space.
\section{Design overview}
\label{sec:design}

Our system is striping-based: a blob is made up of blocks of a fixed size
\emph{psize}, referred to as \emph{pages}. Each page is assigned to a fixed
range of the blob $(k \times psize, (k + 1) \times psize - 1)$. Any range
that covers a full number of pages is said to be \emph{aligned}. These pages 
are distributed among storage space providers. \emph{Metadata} facilitates 
access to a range \emph{(offset, size)} for any existing version of a blob 
snapshot, by associating such a range with the page providers.

A \WRITE or \APPEND generates a \emph{new} set of pages corresponding to the
offset and size requested to be updated. Metadata is then generated
and ``weaved'' together with the old metadata in such way as to
create the illusion of a new incremental snapshot that actually shares
the unmodified pages with the older versions. Thus, two successive snapshots
$v$ and $v + 1$ physically share the pages that fall outside of the range of
the update that generated snapshot $v + 1$.

Consider a read for snapshot $v$ whose range fits exactly a single
page.  The physical page that is accessed was produced by some update
that generated snapshot $w$, with $w \leq v$ such that $w$ is the
highest snapshot version generated by an update whose range intersects
the page. Therefore, when the range of a \READ covers several pages,
these pages may have been generated by different updates. Updates that
do not cover full pages are handled in a slightly more complex way,
but not discussed here, due to space constraints.

\subsection{Architecture overview}
\label{sec:arch}

Our distributed service consists of communicating processes, each fulfilling
a particular role.

\begin{description}

\item[Clients] may create blobs and read, write and append data to
  them. There may be multiple concurrent clients, and their number may
  dynamically vary in time.

\item[Data providers] physically store the pages generated by \WRITE and \APPEND. 
New data providers may dynamically join and leave the system.

\item[The provider manager] keeps information about the available
  storage space.  Each joining provider registers with the provider
  manager. The provider manager decides which providers should be used
  to store the generated pages according to a strategy aiming at
  ensuring an even distribution of pages among providers.

\item[The metadata provider] physically stores the metadata allowing
  clients to find the pages corresponding to the blob snapshot
  version.  Note that the metadata provider may be implemented in a
  distributed way. However, for the sake of readability, we do not
  develop this aspect in our presentation of the algorithms we propose
  for data access. Distributed metadata management is addressed in
  detail in Section~\ref{sec:metadata}.
 
\item[The version manager] is the key actor of the system. It registers 
update requests (\APPEND and \WRITE), assigning snapshot version numbers, end
eventually publishes these updates, guaranteeing total ordering and atomicity.

\end{description}

Our design targets scalability and large-scale distribution.
Therefore, we make a key design choice in avoiding a static role
distribution: any physical node may play one or multiple roles, as a
client, or by hosting data or metadata. This scheme makes our system
suitable for a P2P environment.


\subsection{Reading data}
\label{sec:reads}


The \READ primitive is presented in Algorithm \ref{alg:read}. The
client contacts the version manager first, to check whether the
supplied version \emph{v} has been published and fails if it is not
the case. Otherwise, the client needs find out what pages fully cover
the requested \emph{offset} and \emph{size} for version \emph{v} and
where they are stored. To this purpose, the client contacts the
metadata provider and receives the required metadata. Then it
processes the metadata to generate a set of page descriptors
$PD$. $PD$ holds information about all pages that need to be fetched:
for each page its globally unique page id $pid$, its index $i$ in the
buffer to be read and the page $provider$ that stores it. (Note that,
for the sake of simplicity, we consider here the case where each page
is stored on a single provider. Replication strategies will be
investigated in future work.) Having this information assembled, the
client fetches the pages in parallel and fills the local
\emph{buffer}. Note that the range defined by the supplied
\emph{offset} and $size$ may not be aligned to full pages. In this
case the client may request only a part of the page from the page
provider.

\begin{algorithm}
\caption{\READ}\label{alg:read}
\begin{algorithmic}[1]
\REQUIRE The snapshot version $v$
\REQUIRE The local \emph{buffer} to read to
\REQUIRE The \emph{offset} in the blob 
\REQUIRE The \emph{size} to read
\IF {$v$ is not published}
     \STATE \textbf{fail}
\ENDIF
\STATE $PD \gets$ \emph{READ\_METADATA(v, offset, size)}
\FORALL {$(pid, i, provider) \in PD$ \textbf{in parallel}}
    \STATE read $pid$ from $provider$ into 
    $\mathit{buffer}$ at $i \times psize$
\ENDFOR
\STATE \textbf{return success} 
\end{algorithmic}
\end{algorithm}

Note that, at this stage, for readability reasons, we have not
developed yet the details of metadata management. However, the key
mechanism that enables powerful properties such as efficient
fine-grain access under heavy concurrency relates directly to metadata
management, as discussed in Section~\ref{sec:metadata}.

\subsection{Writing and appending data}
\label{sec:writes}

Algorithm \ref{alg:write} describes how the \WRITE primitive
works. For simplicity, we first consider here aligned writes only,
with page size \emph{psize}. Unaligned writes are also handled by our
system, but, due to space constraints, this case is not discussed
here. The client first needs to determine the number of pages $n$ that
cover the range. Then, it contacts the provider manager requesting a
list of $n$ page providers $PP$ (one for each page) that are capable
of storing the pages.  For each page in parallel, the client generates
a globally unique page id $pid$, contacts the corresponding page
provider and stores the contents of the page on it.  It then updates
the set of page descriptors $PD$ accordingly. This set is later used
to build the metadata associated with this update. After successful
completion of this stage, the client contacts the version manager and
registers its update. The version manager assigns to this update a new
snapshot version $vw$ and communicates it to the client, which then
generates new metadata and ``weaves'' (details in section
\ref{sec:metadata}) it together with the old metadata such that the
new snapshot $vw$ appears as a standalone entity. Finally it notifies
the version manager of success, and returns successfully to the
user. At this point, the version manager takes the responsibility of
eventually publishing $vw$.

\begin{algorithm}
\caption{\WRITE}\label{alg:write}
\begin{algorithmic}[1]
\REQUIRE The local \emph{buffer} used to apply the update.
\REQUIRE The \emph{offset} in the blob.
\REQUIRE The \emph{size} to write.
\ENSURE  The assigned version $vw$ to be published.
\STATE $n \gets (\mathit{offset} + size) / psize$
\STATE $PP \gets$ the list of $n$ page providers
\STATE $PD \gets \emptyset$ 
\FORALL {$0 \leq i < n$ \textbf{in parallel}}
     \STATE $pid \gets$ unique page id
     \STATE $provider \gets PP[i]$
     \STATE store page $pid$ from $\mathit{buffer}$ at $i \times psize$ to $provider$
     \STATE $PD \gets PD \cup {(pid, i, provider)}$
\ENDFOR
\STATE $vw \gets$ assigned snapshot version\label{alg:assign_snapshot}
\STATE \emph{BUILD\_METADATA(vw, offset, size, PD)}
\STATE notify version manager of success
\STATE \textbf{return} $vw$
\end{algorithmic}
\end{algorithm}

\APPEND is almost identical to the \WRITE, with the difference that an
\emph{offset} is directly provided by the version manager at the time
when snapshot version is assigned. This offset is the size of the
previously published snapshot version.

Note that our algorithm enables a high degree of parallelism: for any
update (\WRITE or \APPEND), pages may be asynchronously sent and
stored in parallel on providers. Moreover, multiple clients may
perform such operations with full parallelism: no synchronization is
needed for writing the data, since each update generates new
pages. Some synchronization is necessary when writing the
\emph{metadata}, however the induced overhead is low (see
Section~\ref{sec:metadata}).

\section{Metadata management}
\label{sec:metadata}

\begin{figure*}
  \centerline{%
  \hfill
  \subfigure[The metadata after a write of four pages]%
  {\label{fig:metadata_simple}
    \includegraphics[type=eps,ext=.eps,read=.eps,width=.22\textwidth]%
    {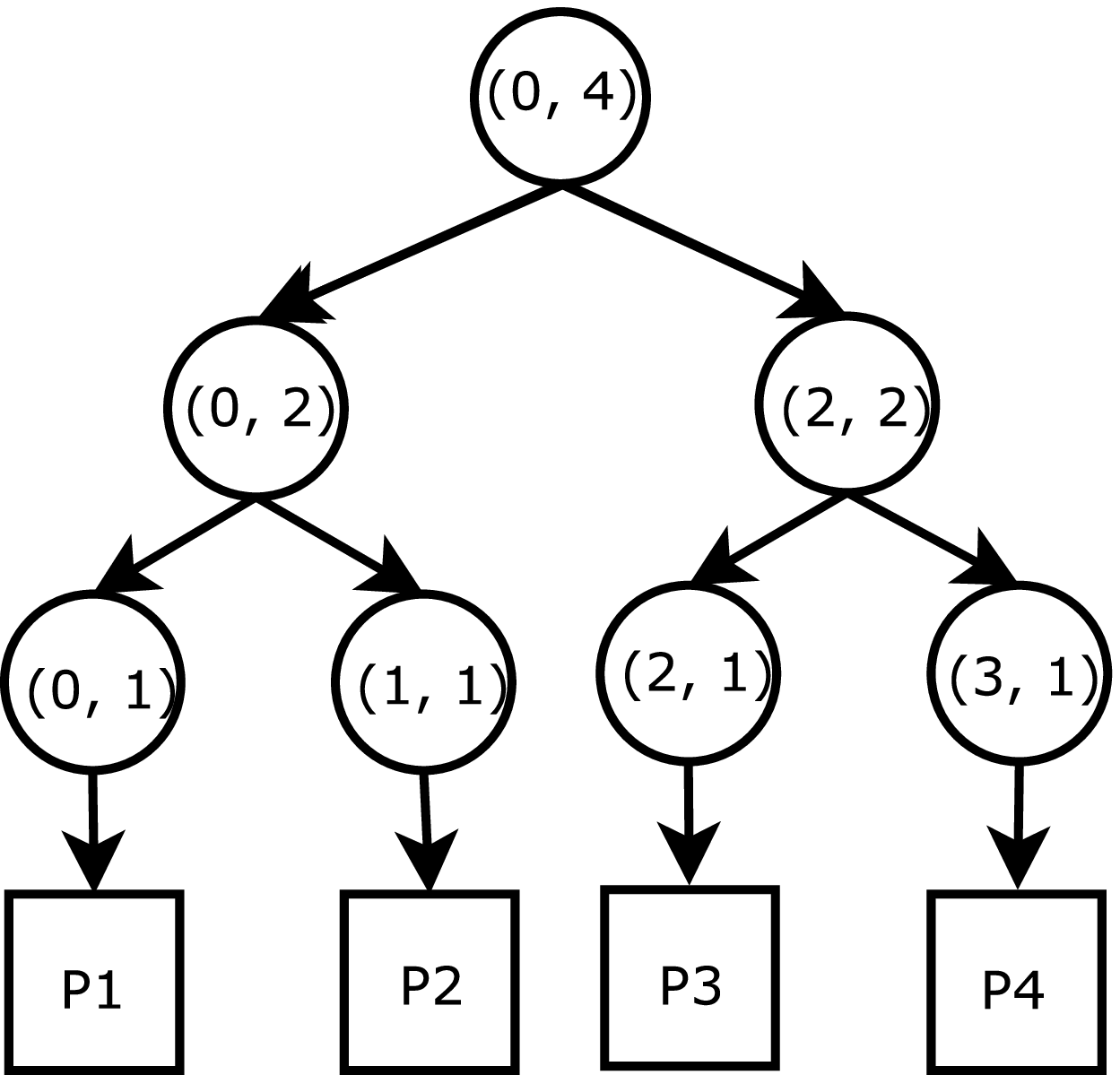}}
  \hfill
  \subfigure[The metadata after overwriting two pages]
  {\label{fig:metadata_inter}
    \includegraphics[type=eps,ext=.eps,read=.eps,width=.28\textwidth]%
    {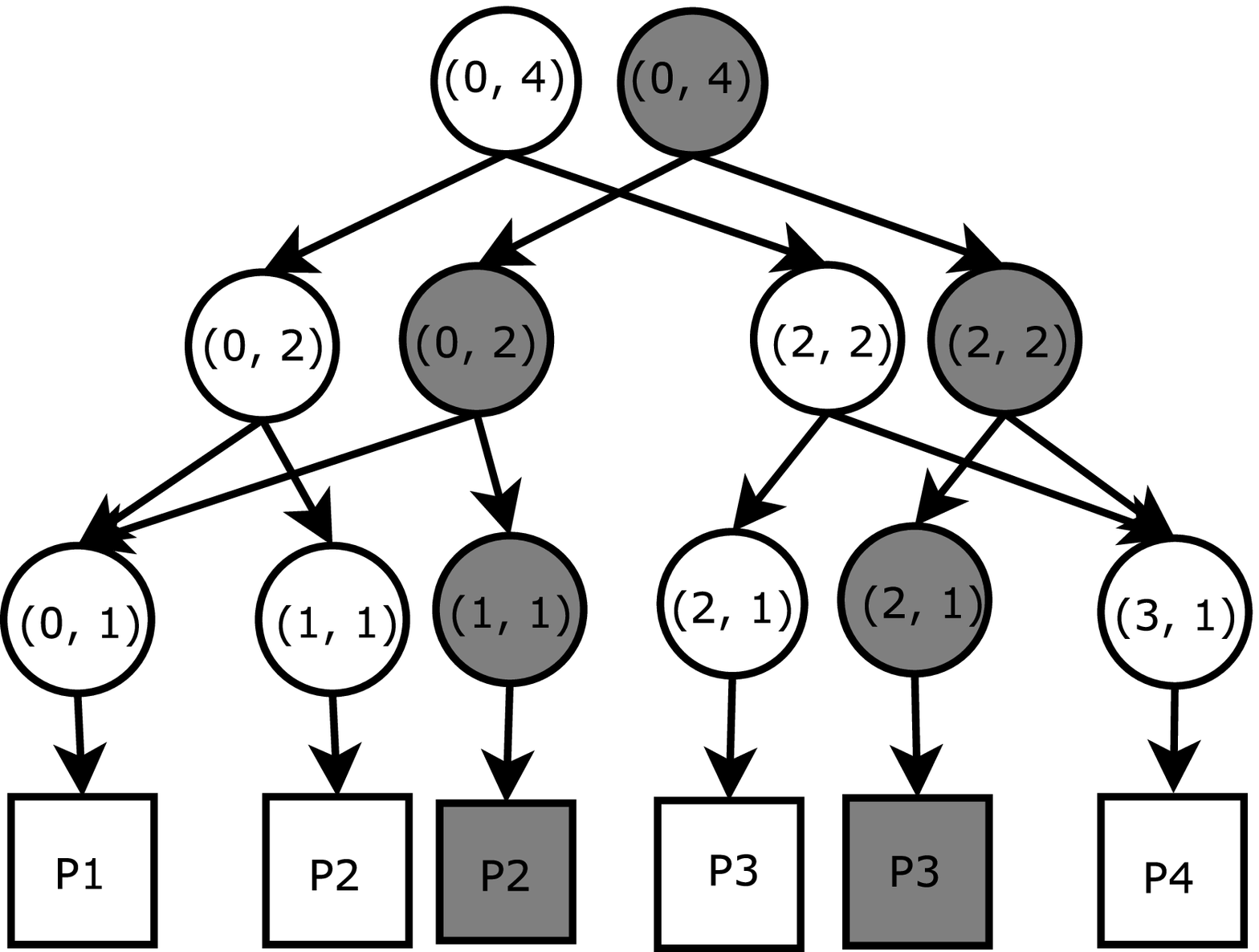}}
  \hfill
  \subfigure[The metadata after an append of one page]
  {\label{fig:metadata_weaved}
    \includegraphics[type=eps,ext=.eps,read=.eps,width=.33\textwidth]%
    {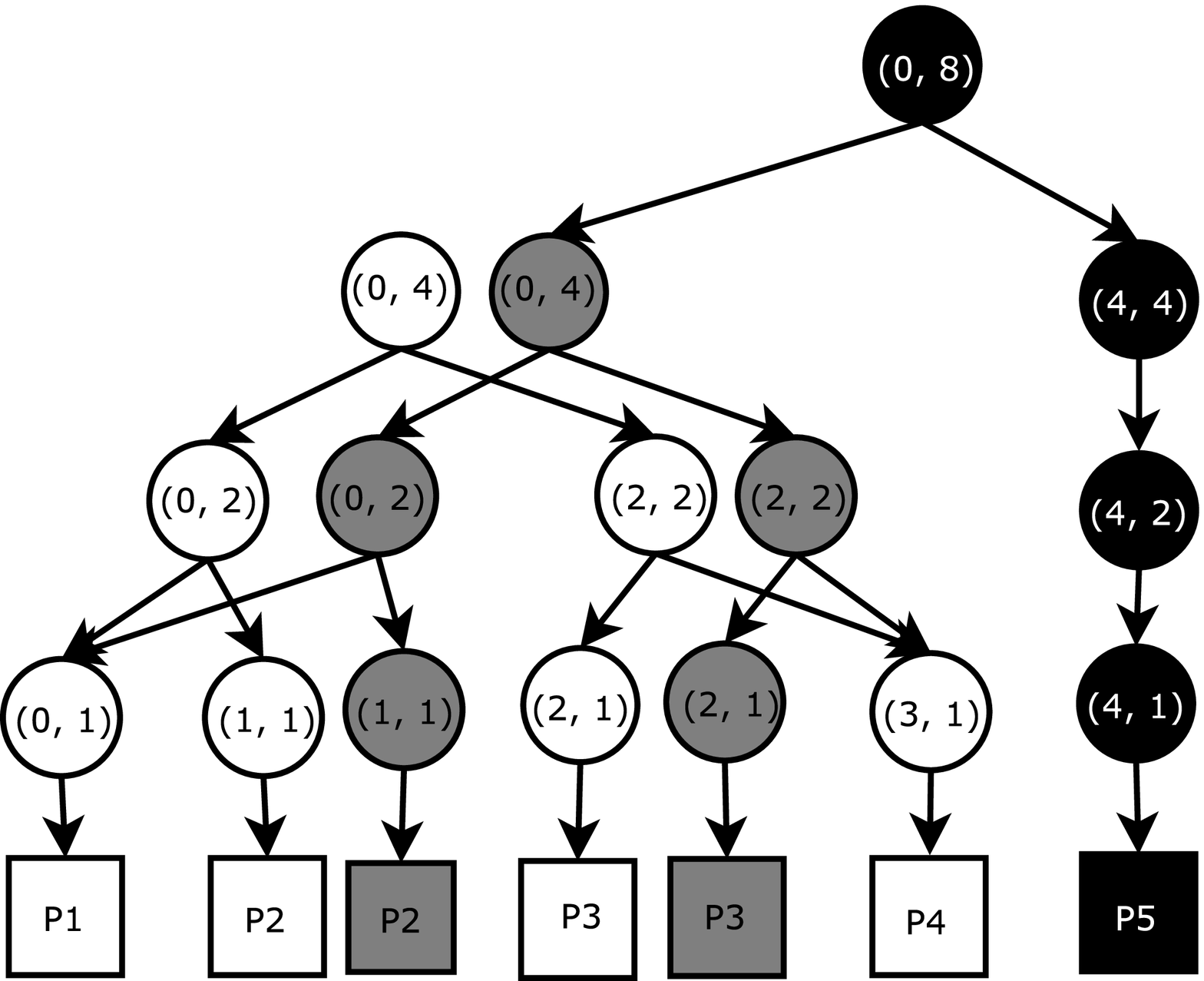}}
  \hfill
  }
  \caption{Metadata representation}%
  \label{fig:metadata}%
\end{figure*}

Metadata stores information about the pages which make up a given
blob, for each generated snapshot version. We choose a simple, yet
versatile design, allowing the system to efficiently build a full view
of the new snapshot of the blob each time an update occurs. This is
made possible through a key design choice: when updating data, new
metadata is created, rather than updating old metadata. As we will
explain below, this decision significantly helps us provide support
for heavy concurrency, as it favors \emph{independent concurrent
  accesses to metadata without synchronization}.

\subsection{The distributed metadata tree}

We organize metadata as a \emph{distributed segment
  tree}~\cite{Zheng06Distributed-Segment-NQ}, one associated to each
snapshot version of a given blob $id$. A segment tree is a binary tree
in which each node is associated to a range of the blob, delimited by
\emph{offset} and \emph{size}. We say that the node \emph{covers} the
range (\emph{offset}, \emph{size}). For each node that is not a leaf,
the left child covers the first half of the range, and the right child
covers the second half. Each leaf covers a single page. We assume the
page size $psize$ is a power of two.

For example, Figure \ref{fig:metadata_simple} depicts the structure of
the metadata for a blob consisting of four pages. We assume the page
size is 1. The root of the tree covers the range $(0, 4)$ (i.e.,
\emph{offset} = 0, \emph{size} = 4~pages), while each leaf covers
exactly one page in this range.

Tree nodes are stored on the metadata provider in a distributed way,
using a simple DHT (Distributed Hash Table). This choice favors
concurrent access to metadata, as explained in
Section~\ref{sec:algo:metadata}. Each tree node is identified
uniquely by its version and range specified by the offset and size it
covers.  Inner nodes hold the version of the left child $vl$ and the
version of the right child $vr$, while leaves hold the page id $pid$
and the $provider$ that store the page.

\paragraph{Sharing metadata across snapshot versions}
Such a metadata tree is created when the first pages of the blob are
written, for the range covered by those pages. Note that rebuilding a
\emph{full} tree for subsequent updates would be space- and
time-inefficient. This can be avoided by \emph{sharing} the existing
tree nodes that cover the blob ranges which do \emph{not} intersect
with the range of the update to be processed. Of course, new tree
nodes are created for the ranges that \emph{do} intersect with the
range of the update. These new tree nodes are ``weaved'' with existing
tree nodes generated by past updates, in order to build a a new
consistent view of the blob, corresponding to a new snapshot
version. This process is illustrated on
Figures~\ref{fig:metadata_simple} and ~\ref{fig:metadata_inter}.
Figure~\ref{fig:metadata_simple} corresponds to an initial snapshot
version (1) of a 4-page blob, whereas Figure~\ref{fig:metadata_inter}
illustrates how metadata evolves when pages 2 and 3 of this blob are
modified (snapshot version 2). Versions are color-coded: the initial
snapshot 1 is white, snapshot 2 is grey.  When a \WRITE updates the
second and third page of the blob, the grey nodes are generated: $(1,
1)$, $(2, 1)$, $(0, 2)$, $(2, 2)$, $(0, 4)$. These new grey nodes are
``weaved'' with existing white nodes corresponding to the unmodified
pages 1 and 4. Therefore, the left child of the grey node that covers
$(0, 2)$ is the white node that covers $(0, 1)$; similarly, the right
child of the grey node that covers $(2, 2)$ is the white node covering
$(3, 1)$.

\paragraph{Expanding the metadata tree}

\APPEND operations make the blob ``grow'': consequently, the metadata
tree gets expanded, as illustrated on
Figure~\ref{fig:metadata_weaved}. Continuing our example, we assume
that the \WRITE generating snapshot version 2 is followed by an
\APPEND for one page, which generates snapshot version 3 of the blob
(black-colored). New metadata tree nodes are generated, to take into
account the creation of the fifth page.  The left child of the new
black root,$(0, 8)$ is the old, grey root of snapshot 2, $(0,
4)$.

\subsection{Accessing metadata: algorithms}
\label{sec:algo:metadata}

\paragraph{Reading metadata}

During a \READ, the metadata is accessed (Algorithm
\ref{alg:read_meta}) in order to find out what pages fully cover the
requested range $R$ delimited by \emph{offset} and $size$. It is
therefore necessary to traverse down the segment tree, starting from
the root that corresponds to the requested snapshot version.  A node
$N$ that covers segment $R_N$ is explored if the intersection of $R_N$
with $R$ is not empty.  All explored leaves reached this way are used
to build the set of page descriptors $PD$ that is used to fetch the
contents of the pages. To simplify the presentation of the algorithm,
we introduce two primitives.  $GET\_NODE(v, \mathit{offset}, size)$
fetches and returns the contents of the node identified by the
supplied version, offset and size from the metadata provider.
Similarly, $GET\_ROOT(v)$ fetches and returns the root of the tree
corresponding to version $v$.

\paragraph{Writing metadata}

\begin{algorithm}
\caption{$READ\_META$}\label{alg:read_meta}
\begin{algorithmic}[1]
\REQUIRE The snapshot version $v$ 
\REQUIRE The $\mathit{offset}$ of the blob 
\REQUIRE The $size$ to read
\ENSURE The set of page descriptors $PD$
\STATE $NS \gets $\{$GET\_ROOT(v)$\}
\WHILE {$NS \neq \emptyset$}
\STATE $N \gets $ extract node from $NS$
\IF{$N$ is leaf}
\STATE $i \gets (\mathit{N.offset} - \mathit{offset}) / psize$
\STATE $PD \gets PD \cup {(N.pid, i, N.provider)}$
\ELSE
\IF {$(\mathit{offset}, size)$ intersects $(N.offset, N.size / 2)$}
\STATE $NS \gets NS \cup GET\_NODE(N.vl, N.offset, N.size / 2)$
\ENDIF
\IF {$(\mathit{offset}, size)$ intersects $(N.offset + N.size / 2, N.size / 2)$}
\STATE $NS \gets NS \cup GET\_NODE(N.vr, N.offset + N.size / 2, N.size / 2)$
\ENDIF
\ENDIF
\ENDWHILE
\end{algorithmic}
\end{algorithm}

For each update (\WRITE or \APPEND) producing snapshot version
\emph{vw}, it is necessary to build a new metadata tree (possibly
sharing nodes with the trees corresponding to previous snapshot
versions). This new tree is the smallest (possibly incomplete) binary
tree such that its leaves are exactly the leaves covering the pages of
range that is written. The tree is built bottom-up: first the leaves
corresponding to the newly generated pages are built, then the inner
nodes $P$ are built up to (and including) the root. This process is
illustrated in Algorithm~\ref{alg:write_meta}. Note that inner nodes
may have children which do \emph{not} intersect the range of the
update to be processed. For any given snapshot version $vw$, these
nodes form the \emph{set of border nodes} $B_{vw}$. When building the
metadata tree, the algorithm needs to compute the corresponding
versions of such children nodes ( $vl$ or $vr$). For simplicity, we do
not develop here how the set of border nodes is computed before
building the tree.

\begin{algorithm}
\caption{$BUILD\_META$}\label{alg:write_meta}
\begin{algorithmic}[1]
\REQUIRE The assigned snapshot version $vw$.
\REQUIRE The $\mathit{offset}$ in the blob.
\REQUIRE The $size$ to write.
\REQUIRE The set of page descriptors $PD$.
\ENSURE     
\STATE $Q \gets \emptyset$
\STATE $B_{vw} \gets$ build the set of border nodes
\FORALL {$(pid, i, provider) \in PD$}
    \STATE $N \gets NEW\_NODE(vw, offset + i \times psize, psize)$
    \STATE $N.pid \gets pid$
    \STATE $N.provider \gets provider$
    \STATE $Q \gets Q \cup \{N\}$
\ENDFOR
\STATE $V \gets Q$
\WHILE {$Q \neq \emptyset$}
    \STATE $N \gets$ extract a node from $Q$
    \IF {$N$ is not root}
        \IF {$N.\mathit{offset} \% (2 \times N.size) = 0$}
            \STATE $P \gets NEW\_NODE(vw, N.\mathit{offset}, 2 \times N.size)$
            \STATE $position \gets LEFT$
        \ELSE
            \STATE $P \gets NEW\_NODE(vw, N.\mathit{offset} - N.size, 2 \times N.size)$
            \STATE $position \gets RIGHT$
        \ENDIF
        \IF {$P$ not $\in V$}
            \IF {$position = LEFT$}
                \STATE $P.vl \gets vw$
                \STATE $P.vr \gets$ extract right child version from $B_{vw}$
            \ENDIF
            \IF {$position = RIGHT$}
               \STATE $P.vr \gets vw$
                \STATE $P.vl \gets$ extract left child version from $B_{vw}$
            \ENDIF
            \STATE $Q \gets Q \cup \{P\}$
            \STATE $V \gets V \cup \{P\}$
        \ENDIF
    \ENDIF
\ENDWHILE
\FORALL {$N \in V$ \textbf{in parallel}}
    \STATE write $N$ to the metadata provider
\ENDFOR
\end{algorithmic}
\end{algorithm}

\paragraph{Why {\WRITE}s and {\APPEND}s may proceed in parallel}

Building new metadata tree nodes might seem to require
serialization. Consider two concurrent clients $C_1$ and $C_2$. Let us
assume that, after having written their pages in parallel, with no
synchronization, that contact the version manager to get their
snapshot versions. Let us assume $C_1$ gets snapshot versions $vw$ and
$C_2$ gets snapshot version $vw + 1$. The two clients should then
start to build their metadata tree nodes concurrently. However, it may
seem that client $C_2$ must wait for client $C_1$ to complete writing
metadata, because tree nodes built by $C_1$ may actually be part of
the set of border nodes of $C_2$, which is used by $C_2$ to build its
own tree nodes.

As our goal is to favor concurrent {\WRITE}s and {\APPEND}s (and,
consequently, concurrent metadata writing), we choose to avoid such a
serialization by introducing a small computation overhead.  Note that
$C_2$ may easily compute the border node set $B_2$ by descending the
tree (starting from the root) corresponding to snapshot $vw$ generated
by $C_1$. It may thus gather all left and right children of the nodes
which intersect the range of the update corresponding to snapshot $vw
+ 1$.  If the root of snapshot $vw + 1$ covers a larger range than the
root of snapshot $vw$, then the set of border nodes contains exactly
one node: the root of snapshot $vw$. Our main difficulty comes from
the nodes that are build by $C_1$ that can actually be part of set of
border nodes of $C_2$, because all other nodes of the set of border
nodes of $C_2$ can be computed as described above, by using the root of
the latest published snapshot $vp$ instead of the root of $vw$.

Our solution to this is to introduce a small computation overhead in
the version manager, who will supply the problematic tree nodes that
are part of the set of border nodes directly to the writer at the
moment it is assigned a new snapshot version. This is possible because
the range of each concurrent \WRITE or \APPEND is registered by the
version manager. Such operations are considered said to be
\emph{concurrent} with the update being processed if they have been
assigned a version number (after writing their data), but they have
not been published yet (e.g. because they have not finished writing
the corresponding metadata). By iterating through the concurrent \WRITE
and \APPEND operations (which have been assigned a lower snapshot
version), the version manager will build the partial set of border
nodes and provide it to the writer when it asks for the snapshot
version. The version manager also supplies a recently published
snapshot version that can be used by the writer to compute the rest of
the border nodes. Armed with both the partial set of border nodes and
a published snapshot version, the writer is now able to compute the
full set of border nodes with respect to the supplied snapshot
version.

\subsection{Discussion}
\label{sec:discussion}

Our approach enables efficient versioning both in terms of performance
under heavy load and in terms of required storage space. Below we
discuss some of the properties of our system.

\paragraph{Support for heavy access concurrency}
 
Given that updates always generate new pages instead of overwriting
older pages, \READ, \WRITE and \APPEND primitives called by concurrent
clients may fully proceed in parallel at the application-level, with
no need for explicit synchronization. This is a key feature of our
distributed algorithm. Internally, synchronization is kept minimal:
distinct pages may read or updated in a fully parallel way; data
access serialization is only necessary when the same provider is
contacted at the same time by different clients, either for reading or
for writing. It is important to note that the strategy employed by the
provider manager for page-to-provider distribution plays a central
role in minimizing such conflicts that lead to serialization.

Note on the other hand that internal serialization is necessary when
two updates (\WRITE or \APPEND) contact the version manager
simultaneously to obtain a snapshot version. This step is hoverer
negligible when compared to the full operation.

Finally, the scheme we use for metadata management also aims at
enabling parallel access to metadata as much as possible. The
situations where synchronization is necessary have been discussed in
Section~\ref{sec:algo:metadata}.

\paragraph{Efficient use of storage space} 

Note that new storage space is necessary for newly written pages only:
for any \WRITE or \APPEND, the pages that are NOT updated are
physically shared by the newly generated snapshot version with the
previously published version. This way, the same physical page may be
shared by a large number of snapshot versions of the same
blob. Moreover, as explained in Section~\ref{sec:metadata}, multiple
snapshot versions may partially share metadata.

\paragraph{Atomicity} 

Recent arguments~\cite{1270853, RJ10316,1024400} stress the need to
provide \emph{atomicity} for operations on objects. An atomic storage
algorithm must guarantee any read or write operation appears to
execute instantaneously between its invocation and completion despite
concurrent access from any number of clients. In our architecture, the
version manager is responsible for assigning snapshot versions and for
publishing them upon successful completion of {\WRITE}s and
{\APPEND}s. Note that concurrent {\WRITE}s and {\APPEND}s work in
complete isolation, as they do not \emph{modify}, but rather
\emph{add} data and metadata. It is then up the the version manager to
decide when their effects will be revealed to the other clients, by
publishing their assigned versions in a consistent way.  The only
synchronization occurs at the level at the version manager. In our
current implementation, atomicity is easy to achieve, as the version
manager is centralized. Using a distributed version manager will be
addressed in the near future.

%

\section{Experimental evaluation}

We experimented and evaluated the approach developed above within the
framework of our BlobSeer prototype. To implement the metadata
provider in a distributed way, we have developed a custom DHT
(Distributed Hash Table), based on simple static distribution
scheme. This allows metadata to be efficiently stored and retrieved in
parallel. 


Evaluations have been performed on the Grid'5000~\cite{CapCarDayetal05SC05} testbed,
an experimental Grid platform gathering 9 sites geographically distributed 
in France. In each experiment, we used at most 175 nodes of the Rennes site 
of Grid'5000. Nodes are outfitted with x86\_64 CPUs and 4~GB of RAM. Intracluster 
bandwidth is 1 Gbit/s (measured: 117.5MB/s for TCP sockets with MTU = 1500~B), 
latency is 0.1~ms.

\begin{figure*}
  \centerline{%
  \subfigure[Append throughput as a blob dynamically grows]%
  {\label{fig:append_simple}
    \includegraphics[type=eps,ext=.eps,read=.eps,width=.45\textwidth]%
    {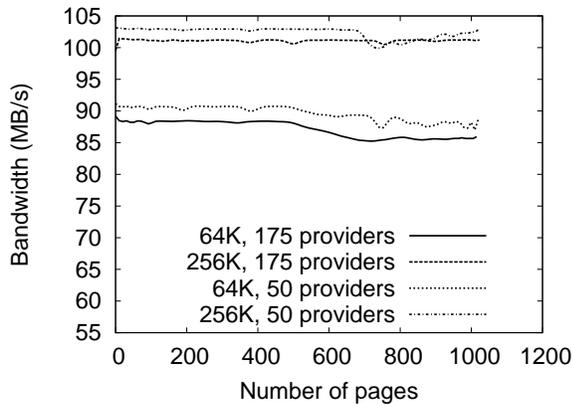}}
  \hfill
  \subfigure[Read throughput under concurrency]%
  {\label{fig:multi_reads}
    \includegraphics[type=eps,ext=.eps,read=.eps,width=.45\textwidth]%
    {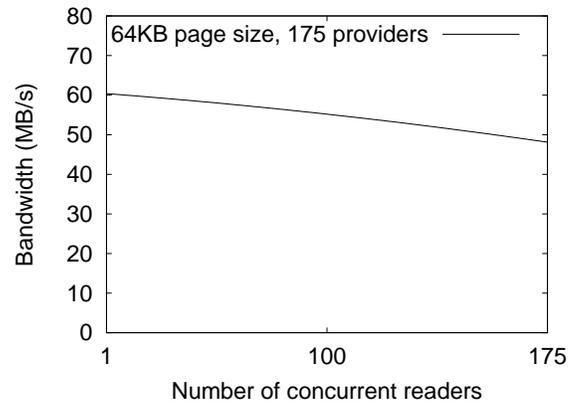}}
  }
  \caption{Experimental results: data throughput}%
  \label{fig:results}%
\end{figure*}


We first ran a set of experiments to evaluate the impact of our
metadata scheme over the performance of the \APPEND operation (in
terms of bandwidth), while the blob size continuously grows. In these
experiments, a single client process creates an empty blob and starts
appending to it, while we constantly monitor the bandwidth of the
\APPEND operation.

We use two deployment settings. We deploy each the version manager and
the provider manager on two distinct dedicated nodes, and we co-deploy
a data provider and a metadata provider on the other nodes,
using a total of 50 data and metadata providers in the first setting
and 175 data and metadata providers in the second setting. In each of
the two settings, a client creates a blob and starts appending 64~MB
of data to the blob. This process is repeated two times, using a
different page size each time: 64~KB and 256~KB.

Results are shown in Figure \ref{fig:append_simple}: they show that a
high bandwidth is maintained even when the blob grows to large sizes,
thus demonstrating a low metadata overhead. A slight bandwidth
decrease is observed when the number of pages reaches a power of two:
this corresponds to the expected metadata overhead increase caused by
adding a new level to the metadata tree.
%

A second set of experiments evaluates the bandwidth performance when
multiple readers access disjoint parts of the same blob. We use again 175
nodes for these experiments. As in the previous
case, the version manager and the provider manager are deployed on two
dedicated nodes, while a data provider and a metadata provider are
co-deployed on the remaining 173 nodes.

In a first phase, a single client appends data to the blob until the
blob grows to 64~GB.  Then, we start reading the first 64~MB from the
blob with a single client. This process is repeated 100 times and the
average bandwidth computed. Then, we increase the number of readers
to 100. The readers are deployed on nodes that already run a data and
metadata provider. They concurrently read distinct 64~MB chunks from
the blob.  Again, the process in repeated 100 times, and the average
read bandwidth is computed.  In a last step the number of concurrent
clients is increased to 175 and the same process repeated, obtaining
another average read bandwidth.

Note that, even there is no conflict for accessing the same data, the
readers concurrently traverse the metadata tree, whose nodes may be
concurrently requested by multiple readers. Note also that, as the
number of pages of the blob is very large with respect to the total
number of available metadata and data providers, each physical node
may be subject to heavily concurrent requests.

The obtained results are represented in Figure \ref{fig:multi_reads},
where the average read bandwidth is represented as a function of the
number of concurrent readers, interpolated to fit the three
experiments. We observe a very good scalability: the read bandwidth
drops from 60MB/s for a single reader to 49MB/s for 175 concurrent
readers. 


\label{sec:impl}

\section{Conclusion}
\label{sec:conclusion}

As more and more application classes and services start using the P2P
paradigm in order to achieve high scalability, the demand for
adequate, scalable data management strategies is ever higher. One
important requirement in this context is the ability to efficiently
cope with accesses to continuously growing data, while supporting a
highly concurrent, highly distributed environment. We address this
requirement for the case of huge \emph{unstructured} data. We propose
an efficient versioning scheme allowing a large number of clients to
concurrently \emph{read, write and append} data to binary large
objects (blobs) that are fragmented and distributed at a very large
scale.  Our algorithms guarantees atomicity while still achieving a
good data access performance. To favor scalability under heavy
concurrency, we rely on an original metadata scheme, based on a
distributed segment tree that we build on top of a Distributed Hash
Table (DHT). Rather than modifying data and metadata in place when
data updates are requested, new data fragments are created, and the
corresponding metadata are ``weaved'' with the old metadata, in order
to provide a new view of the whole blob, in a space-efficient
way. This approach favors independent, concurrent accesses to data and
metadata without synchronization and thereby enables a high throughput
under heavy concurrency. The proposed algorithms have been implemented
within our \emph{BlobSeer} prototype and experimented on the Grid'5000
testbed, using up to 175 nodes: the preliminary results suggest a good
scalability with respect to the data size and to the number of
concurrent accesses. Further experimentations are in progress, which
aim at demonstrating the benefits of data and metadata
distribution. We also intend to investigate extensions to
our approach allowing to add support for volatility and failures.

\bibliographystyle{abbrv}
\bibliography{juxmem,projects,blobseer}

\end{document}